\documentstyle[prd,aps,preprint,tighten,epsfig]{revtex}

\begin{document}

\draft

\title{Low-energy limits on heavy Majorana neutrino masses from the
neutrinoless double-beta decay and non-unitary neutrino mixing}
\author{{\bf Zhi-zhong Xing}
\thanks{E-mail: xingzz@ihep.ac.cn}}
\address{Institute of High Energy Physics
and Theoretical Physics Center for Science Facilities, \\ Chinese
Academy of Sciences, P.O. Box 918, Beijing 100049, China}
\maketitle

\begin{abstract}
In the type-I seesaw mechanism, both the light Majorana neutrinos
($\nu^{}_1, \nu^{}_2, \nu^{}_3$) and the heavy Majorana neutrinos
($N^{}_1, \cdots, N^{}_n$) can mediate the neutrinoless
double-beta ($0\nu\beta\beta$) decay. We point out that the
contribution of $\nu^{}_i$ to this $0\nu\beta\beta$ process is
also dependent on the masses $M^{}_k$ and the mixing parameters
$R^{}_{ek}$ of $N^{}_k$ as a direct consequence of the exact
seesaw relation, and the effective mass term of $\nu^{}_i$ is in
most cases dominant over that of $N^{}_i$. We obtain a new bound
$|\sum R^2_{ek} M^{}_k| < 0.23 ~ {\rm eV}$ (or $< 0.85 ~ {\rm eV}$
as a more conservative limit) at the $2\sigma$ level, which is
much stronger than $|\sum R^2_{ek} M^{-1}_k| < 5 \times 10^{-8} ~
{\rm GeV}^{-1}$ used in some literature, from current experimental
constraints on the $0\nu\beta\beta$ decay. Taking the minimal
type-I seesaw scenario for example, we illustrate the possibility
of determining or constraining two heavy Majorana neutrino masses
by using more accurate low-energy data on lepton number violation
and non-unitarity of neutrino mixing.
\end{abstract}


\newpage

\section{Introduction}

Almost undebatable evidence for finite neutrino masses and large
neutrino mixing angles has recently been achieved from solar,
atmospheric, reactor and accelerator neutrino oscillation
experiments \cite{SNO,SK,KM,K2K} . This exciting breakthrough
opens a new window to physics beyond the standard model (SM),
since the SM itself only contains three massless neutrinos (i.e.,
$\nu^{}_e$, $\nu^{}_\mu$ and $\nu^{}_\tau$, corresponding to the
mass eigenstates $\nu^{}_1$, $\nu^{}_2$ and $\nu^{}_3$). The
simplest way to generate non-zero but tiny neutrino masses
$m^{}_i$ for $\nu^{}_i$ is to extend the SM by introducing at
least two right-handed neutrinos and allowing lepton number
violation. In this well-known (type-I) seesaw mechanism
\cite{SS1}, the $SU(2)_{\rm L} \times U(1)_{\rm Y}$
gauge-invariant mass terms of charged leptons and neutrinos are
given by
\begin{equation}
-{\cal L}^{}_{\rm mass} \; =\; \overline{l^{}_{\rm L}} Y^{}_l H
E^{}_{\rm R} + \overline{l^{}_{\rm L}} Y^{}_\nu \tilde{H}
N^{}_{\rm R} + \frac{1}{2} \overline{N^{c}_{\rm R}} M^{}_{\rm R}
N^{}_{\rm R} + {\rm h.c.} \; ,
%     (1)
\end{equation}
where $\tilde{H} \equiv i\sigma^{~}_2 H^*$, $l_{\rm L}$ denotes
the left-handed lepton doublet, and $M^{}_{\rm R}$ is the mass
matrix of right-handed neutrinos. After the spontaneous gauge
symmetry breaking, we arrive at the charged-lepton mass matrix
$M^{}_l = Y^{}_l v$ and the Dirac neutrino mass matrix $M^{}_{\rm
D} = Y^{}_\nu v$, where $v \simeq 174 ~ {\rm GeV}$ is the vacuum
expectation value of the neutral component of the Higgs doublet
$H$ and characterizes the Fermi scale of weak interactions. The
mass scale of $M^{}_{\rm R}$ (or equivalently the seesaw scale
$\Lambda^{}_{\rm SS}$) is crucial, because it is relevant to
whether the seesaw mechanism itself is theoretically natural and
experimentally testable. Although $\Lambda^{}_{\rm SS} \ll v$ is
not impossible \cite{DG}, it is in general expected that
$\Lambda^{}_{\rm SS}$ should be much higher than the Fermi scale.
In particular, the conventional seesaw mechanism works at a scale
which is not far away from the scale of grand unified theories.
Driven by the upcoming running of the Large hadron Collider (LHC),
more and more attention has been paid to the TeV scale at which
the unnatural gauge hierarchy problem of the SM may be solved or
softened by new physics. If the TeV scale is really a fundamental
scale, then we are reasonably motivated to speculate that possible
new physics existing at this scale and responsible for the
electroweak symmetry breaking might also be responsible for the
origin of neutrino masses. In this sense, it is meaningful to
investigate the TeV seesaw mechanism and balance its theoretical
naturalness and experimental testability at the energy frontier
set by the LHC \cite{ICHEP08}.

\vspace{0.4cm}

A direct test of the type-I seesaw mechanism demands the
experimental discovery of heavy Majorana neutrinos $N^{}_k$ (for
$k=1, \cdots, n$) at the LHC, but two prerequisites must be
satisfied: (a) their masses $M^{}_k$ must be of ${\cal O}(1)$ TeV
or smaller; and (b) their couplings to charged leptons
$R^{}_{\alpha k}$ (for $\alpha = e, \mu, \tau$ and $k=1, \cdots,
n$) must not be too small. The strongest bound on $M^{}_k$ and
$R^{}_{e k}$ comes from the non-observation of the neutrinoless
double-beta ($0\nu\beta\beta$) decay \cite{Review}, as $N^{}_k$
can mediate this lepton-number-violating process. Current
experimental lower limit on the half-lifetime of the
$0\nu\beta\beta$ decay is usually translated into
\begin{equation}
\left| \sum^n_{k=1} \frac{R^2_{e k}}{M^{}_k} \right| \; < \; 5
\times 10^{-8} ~ {\rm GeV}^{-1} \;
%     (2)
\end{equation}
in some literature \cite{Seesaw}. In obtaining Eq. (2), one has
ignored the contribution of three light Majorana neutrinos
$\nu^{}_i$ (for $i=1,2,3$) to the $0\nu\beta\beta$ decay.

\vspace{0.4cm}

The first purpose of this paper is to point out that the
constraint in Eq. (2) is not always useful for the type-I seesaw
mechanism either at a superhigh-energy scale or at the electroweak
or TeV scale. The reason is simply that the contribution of
$\nu^{}_i$ to the $0\nu\beta\beta$ decay is in most cases dominant
over the contribution of $N^{}_k$ to the same process, leading to
a much stronger bound on $M^{}_k$ and $R^{}_{ek}$ through the
exact seesaw relation:
\begin{equation}
\left| \sum^n_{k=1} R^2_{e k} M^{}_k \right| \; < \;  0.23 ~ {\rm
eV} ~~~ ({\rm or} ~ < 0.85 ~ {\rm eV}) \;
%     (3)
\end{equation}
at the $2\sigma$ level, which is equivalent to $\langle
m\rangle^{}_{ee} < 0.23 ~ {\rm eV}$ (or $< 0.85$ eV as a more
conservative bound) \cite{CC,Fogli} for the effective mass of the
$0\nu\beta\beta$ decay mediated by $\nu^{}_i$. The second purpose
of this paper is to look at whether the future measurements of
lepton number violation and non-unitarity of neutrino flavor
mixing are possible to shed light on $M^{}_k$. Taking the minimal
type-I seesaw scenario \cite{MSS} for example, we shall illustrate
the possibility of determining or constraining two heavy Majorana
neutrino masses by using more accurate low-energy data on the
$0\nu\beta\beta$ decay and non-unitary neutrino mixing and CP
violation.

\section{Stronger bound on the $0\nu\beta\beta$ decay}

After the spontaneous gauge symmetry breaking (i.e.,
$SU(2)^{}_{\rm L} \times U(1)^{}_{\rm Y} \rightarrow U(1)^{}_{\rm
em}$), the mass terms in Eq. (1) turn out to be
\begin{eqnarray}
-{\cal L}^\prime_{\rm mass} \; = \; \overline{E^{~}_{\rm L}}
M^{~}_l E^{~}_{\rm R} + \frac{1}{2} ~ \overline{\left( \nu^{}_{\rm
L} ~N^c_{\rm R}\right)} ~ \left( \matrix{ {\bf 0} & M^{}_{\rm D}
\cr M^T_{\rm D} & M^{}_{\rm R}}\right) \left( \matrix{ \nu^c_{\rm
L} \cr N^{}_{\rm R}}\right) + {\rm h.c.} \; ,
%     (4)
\end{eqnarray}
where $E$ and $\nu^{~}_{\rm L}$ stand respectively for the column
vectors of $(e, \mu, \tau)$ and $(\nu^{~}_e, \nu^{~}_\mu,
\nu^{~}_\tau)^{~}_{\rm L}$. Without loss of generality, one may
take $M^{}_l = {\rm Diag}\{m^{}_e, m^{}_\mu, m^{}_\tau\}$. The
overall $(3+n) \times (3+n)$ neutrino mass matrix in Eq. (4) can
be diagonalized by a unitary transformation; i.e.,
\begin{eqnarray}
\left( \matrix{ {\bf 0} & M^{}_{\rm D} \cr M^T_{\rm D} & M^{}_{\rm
R}}\right) = \left(\matrix{V & R \cr S & U}\right) \left( \matrix{
\widehat{M}^{}_\nu & {\bf 0} \cr {\bf 0} &
\widehat{M}^{}_N}\right) \left(\matrix{V & R \cr S & U}\right)^T
\; ,
%     (5)
\end{eqnarray}
where $\widehat{M}^{}_\nu = {\rm Diag}\{m^{}_1, m^{}_2, m^{}_3\}$
and $\widehat{M}^{}_N = {\rm Diag}\{M^{}_1, \cdots, M^{}_3\}$.
After this diagonalization, the flavor states of three light
neutrinos ($\nu^{}_e, \nu^{}_\mu, \nu^{}_\tau$) can be expressed
in terms of the $(3+n)$ mass states of light and heavy neutrinos
($\nu^{}_1, \nu^{}_2, \nu^{}_3$ and $N^{}_1, \cdots, N^{}_n$), and
thus the standard charged-current interactions between
$\nu^{}_\alpha$ and $\alpha$ (for $\alpha = e, \mu, \tau$) can be
written as
\begin{eqnarray}
-{\cal L}^{}_{\rm cc} \; = \; \frac{g}{\sqrt{2}} ~
\overline{\left(e~~ \mu~~ \tau\right)^{}_{\rm L}} ~\gamma^\mu
\left[ V \left( \matrix{\nu^{}_1 \cr \nu^{}_2 \cr \nu^{}_3}
\right)^{}_{\rm L} + R \left( \matrix{N^{}_1 \cr \vdots \cr
N^{}_n} \right)^{}_{\rm L} \right] W^-_\mu + {\rm h.c.} \;
%     (6)
\end{eqnarray}
in the basis of mass states. So $V$ is just the neutrino mixing
matrix responsible for neutrino oscillations, while $R$ describes
the strength of charged-current interactions between $(e, \mu,
\tau)$ and $(N^{}_1, \cdots, N^{}_n)$. $V$ and $R$ are correlated
with each other through $VV^\dagger + RR^\dagger = {\bf 1}$. Hence
$V$ itself is not exactly unitary in the type-I seesaw mechanism
and its deviation from unitarity is simply characterized by
non-vanishing $R$ \cite{Xing08}.

\vspace{0.4cm}

Note that $V$ and $R$ are also correlated with each other through
the {\it exact} seesaw relation
\begin{equation}
V \widehat{M}^{}_\nu V^T + R \widehat{M}^{}_N R^T \; = \; {\bf 0}
\; , ~~~
%     (7)
\end{equation}
which can directly be derived from Eq. (5). Taking the
($ee$)-elements for both terms on the left-hand side of Eq. (7),
we immediately arrive at
\begin{equation}
\left(V \widehat{M}^{}_\nu V^T\right)^{}_{ee} \; = \; \sum^3_{i=1}
V^2_{ei} m^{}_i \; = \; - \left(R \widehat{M}^{}_N
R^T\right)^{}_{ee} \; = \; - \sum^n_{k=1} R^2_{ek} M^{}_k \; .
%     (8)
\end{equation}
This simple but interesting result implies that the effective mass
of three light Majorana neutrinos in the $0\nu\beta\beta$ decay is
directly associated with the masses, mixing angles and
CP-violating phases of heavy Majorana neutrinos in the type-I
seesaw mechanism:
\begin{equation}
\langle m\rangle^{}_{ee} \; \equiv \; \left|\sum^3_{i=1} V^2_{ei}
m^{}_i\right| \; = \; \left|\sum^n_{k=1} R^2_{ek} M^{}_k\right| \;
.
%     (9)
\end{equation}
Note that both light Majorana neutrinos $\nu^{}_i$ and heavy
Majorana neutrinos $N^{}_k$ can mediate the $0\nu\beta\beta$
decay, as shown in Fig. 1. When the contribution of $N^{}_k$ is
{\it least} suppressed \cite{Referee}, the overall decay width of
the $0\nu\beta\beta$ process in the type-I seesaw scenario can
approximately be expressed as
%%%%%%%%%%%%%%%%%%%%%%%%%%%%%
\footnote{Current calculations of the nuclear matrix elements of
the $0\nu \beta\beta$ decay are quite uncertain due to our poor
knowledge of the nucleon wave functions \cite{Review}. For
point-like nucleons, a Yukawa-type potential has been used to
estimate the nuclear matrix elements \cite{Referee}. The relevant
result remains true even if the finite size of the nucleons is
taken into account \cite{Referee}. We thank the referee for
calling our attention to these points.}
%%%%%%%%%%%%%%%%%%%%%%%%%%%%%
\begin{eqnarray}
\Gamma^{}_{0\nu\beta\beta} & \propto & \left| \sum^3_{i=1}
V^2_{ei} m^{}_i ~ - ~ \sum^n_{k=1} \frac{R^2_{ek}}{M^{}_k} M^2_A
{\cal F}(A, M^{}_k) \right|^2 \nonumber \\
& = & \left| \sum^n_{k=1} R^2_{ek} M^{}_k \left[ 1 +
\frac{M^2_A}{M^2_k} {\cal F}(A, M^{}_k) \right] \right| \; ,
%     (10)
\end{eqnarray}
where $A$ is the atomic number, ${\cal F}(A, M^{}_k) \simeq 0.1$
depending mildly on the decaying nucleus, and $M^{}_A \simeq 900$
MeV \cite{Referee}. Given $M^{}_k \gtrsim 10^2$ GeV, the second
term in the square brackets of Eq. (10) turns out to be $\lesssim
8.1 \times 10^{-6}$. Hence this term is negligible in most cases,
unless the contribution of $\nu^{}_i$ is vanishing or vanishingly
small due to a contrived cancellation among three different
$V^2_{ei} m^{}_i$ terms (or equivalently, among $n$ different
$R^2_{ek} M^{}_k$ terms), which is in principle not impossible.
Let us consider two limits in which the contributions of light and
heavy Majorana neutrinos to $\Gamma^{}_{0\nu\beta\beta}$ are
decoupled.
\begin{itemize}
\item     In the limit of
$\displaystyle\sum^n_{k=1} R^2_{ek} M^{-1}_k {\cal F}(A,M^{}_k)
\rightarrow 0$, which is almost a realistic case, Eq. (10) is
directly simplified to
\begin{equation}
\sqrt{\Gamma^{}_{0\nu\beta\beta}} \;\; \propto \; \langle
m\rangle^{}_{ee} \; =\; \left|\sum^n_{k=1} R^2_{ek} M^{}_k\right|
\; .
%     (11)
\end{equation}
Current experimental data on the $0\nu\beta\beta$ decay yield an
upper bound on this effective mass term, $\langle m\rangle^{}_{ee}
< 0.23 ~{\rm eV}$ at the $2\sigma$ level \cite{Fogli}, which has
extensively been used to constrain the masses, flavor mixing
angles and Majorana CP-violating phases of three light neutrinos
in the unitary limit of $V$. However, it should be kept in mind
that this upper bound corresponds to some ``favorable" values of
the relevant nuclear matrix elements \cite{Review}. If their
``unfavorable" values are used, one may also arrive at $\langle
m\rangle^{}_{ee} < 0.85$ eV at the $2\sigma$ level \cite{Fogli}.

\item     In the limit of
$\displaystyle\sum^n_{k=1} R^2_{ek} M^{}_k \rightarrow 0$, which
is rather contrived, Eq. (10) can be simplified to
\begin{equation}
\sqrt{\Gamma^{}_{0\nu\beta\beta}} \;\; \propto \; \langle
m\rangle^\prime_{ee} \; \equiv \; M^2_A \left|\sum^n_{k=1}
\frac{R^2_{ek}}{M^{}_k} {\cal F}(A, M^{}_k) \right| \;
%     (12)
\end{equation}
as a rough approximation. Imposing the bound $\langle
m\rangle^\prime_{ee} < 0.23 ~ {\rm eV}$ and inputting $M^{}_A
\simeq 900$ MeV and ${\cal F}(A,M^{}_k) \simeq 0.1$
\cite{Referee}, for example, we obtain
\begin{equation}
\left| \sum^n_{k=1} \frac{R^2_{e k}}{M^{}_k} \right| \; < \; 2.8
\times 10^{-9} ~ {\rm GeV}^{-1} \; ,
%     (13)
\end{equation}
which is a bit stronger than the upper bound shown in Eq. (2). If
the more conservative bound $\langle m\rangle^\prime_{ee} < 0.85 ~
{\rm eV}$ is taken, one will arrive at $\sum R^2_{ek} M^{-1}_k <
1.0 \times 10^{-8} ~ {\rm GeV}^{-1}$, much closer to the result
given in Eq. (2). Such rough bounds have been used by a number of
authors in their preliminary studies of possible collider
signatures of heavy Majorana neutrinos \cite{Seesaw}.
\end{itemize}
Note again that we have ignored the mild dependence of ${\cal
F}(A, M^{}_k)$ on the decaying nuclei in the above discussions.
Otherwise, different $0\nu \beta\beta$ decays should be separately
analyzed.

\vspace{0.4cm}

Below Eq. (10), we have pointed out that the contribution of three
light Majorana neutrinos $\nu^{}_i$ to $\Gamma^{}_{0\nu
\beta\beta}$ is dominant in most cases. This observation is
especially true for the conventional type-I seesaw mechanism with
superhigh $M^{}_k$ (e.g., $\max{(M^{}_k)} \sim 10^{15}$ GeV) and
extremely small $R^{}_{\alpha k}$ (e.g., $|R^{}_{\alpha k}| \sim
10^{-13}$) \cite{SSReview}. When the seesaw mechanism is realized
at the electroweak or TeV scale to generate experimentally
accessible signatures of heavy Majorana neutrinos $N^{}_k$ at the
LHC, however, one usually has to require $\max{(M^{}_k)} \lesssim
{\cal O}(1)$ TeV and $|R^{}_{\alpha k}| \gtrsim 10^{-3}$ up to
${\cal O}(0.1)$ \cite{ICHEP08}, which imply a terrible
cancellation in every term of Eq. (8) so as to give rise to tiny
masses of $\nu^{}_i$. Because such a terrible cancellation in
$\langle m\rangle^{}_{ee}$ does not necessarily mean the same
cancellation in $\langle m\rangle^\prime_{ee}$, it is possible to
get $\langle m\rangle^\prime_{ee} \gg \langle m\rangle^{}_{ee}$ as
a special case, as already discussed in Eqs. (12) and (13). But
the situation might become quite subtle if the masses of heavy
Majorana neutrinos are degenerate \cite{London}. Since the
function ${\cal F}(A, M^{}_k)$ depends both on the atomic number
$A$ and the heavy Majorana neutrino masses $M^{}_k$, $\langle
m\rangle^\prime_{ee}$ might be exceedingly small for one decaying
nucleus in the $\langle m\rangle^{}_{ee} \rightarrow 0$ limit but
not for another in the same limit. A careful analysis of the
relative magnitudes of $\langle m\rangle^{}_{ee}$ and $\langle
m\rangle^\prime_{ee}$ for different $0\nu \beta\beta$ decays is
nevertheless beyond the scope of the present paper and will be
done elsewhere.

\vspace{0.4cm}

For $n=3$, $R$ can be parametrized in terms of nine rotation
angles $\theta^{}_{ij}$ and nine phase angles $\delta^{}_{ij}$
(for $i=1,2,3$ and $j=4,5,6$) \cite{Xing08}. In this
representation,
\begin{eqnarray}
\langle m\rangle^{}_{ee} & \;=\; & \left| M^{}_1 s^2_{14} c^2_{15}
c^2_{16} e^{-2i\delta^{}_{14}} + M^{}_2 s^2_{15} c^2_{16}
e^{-2i\delta^{}_{15}} + M^{}_3 s^2_{16} e^{-2i\delta^{}_{16}}
\right| \nonumber \\
& \;\approx\; & \left| M^{}_1 s^2_{14} + M^{}_2 s^2_{15} e^{2i
(\delta^{}_{14} - \delta^{}_{15})} + M^{}_3 s^2_{16} e^{2i
(\delta^{}_{14} - \delta^{}_{16})} \right| \; ,
%     (14)
\end{eqnarray}
where $s^{}_{ij} \equiv \sin\theta^{}_{ij}$ and $c^{}_{ij} \equiv
\cos\theta^{}_{ij}$. The approximation made in Eq. (14) is very
reasonable because $|RR^\dagger|$ is at most of ${\cal
O}(10^{-2})$ \cite{Antusch} and thus all the mixing angles of $R$
must be very small (at most at the ${\cal O}(0.1)$ level). Given
$\langle m\rangle^{}_{ee} \lesssim 1$ eV and $M^{}_1 \approx
M^{}_2 \approx M^{}_3 \sim v$ or ${\cal O}(1)$ TeV, for instance,
the constraint in Eq. (14) implies that two phase differences
$\delta^{}_{14} - \delta^{}_{15}$ and $\delta^{}_{14} -
\delta^{}_{16}$ should be very close to $\pm \pi/2$ in order to
assure significant cancellations among three terms. Fixing $M^{}_i
\sim 10^2$ GeV (for $i=1,2,3$) as an example, we find that the
level of fine-tuning is at least of ${\cal O} (10^{-9})$ with
$s^2_{1j} \sim 10^{-2}$ or ${\cal O} (10^{-7})$ with $s^2_{1j}
\sim 10^{-4}$ (for $j=4,5,6$). Such unnatural cancellations seem
to be unavoidable in the type-I seesaw models at the electroweak
or TeV scale, unless the relevant mixing angles are extremely
small. Current experimental data can only provide us with a rough
bound $s^2_{14} + s^2_{15} + s^2_{16} \lesssim 1.1 \times 10^{-2}$
\cite{Antusch}, unfortunately. In a specific type-I seesaw model
with the fine-tuning conditions $(M^{}_{\rm D})^{}_{3i} \propto
(M^{}_{\rm D})^{}_{2i} \propto (M^{}_{\rm D})^{}_{1i}$ (for
$i=1,2,3$) and $(M^{}_{\rm D})^2_{11}/M^{}_1 + (M^{}_{\rm
D})^2_{12}/M^{}_2 + (M^{}_{\rm D})^2_{13}/M^{}_3 =0$
\cite{Minkowski}, one may obtain $M^{}_\nu = -M^{}_{\rm D}
M^{}_{\rm R} M^T_{\rm D} = {\bf 0}$, which in turn leads to
$\langle m\rangle^{}_{ee} =0$. Then non-zero but tiny $M^{}_\nu$
and $\langle m\rangle^{}_{ee}$ can be achieved by introducing a
small perturbation to the texture of $M^{}_{\rm D}$. If the
magnitudes of $M^{}_k$ are too big or those of $R^{}_{\alpha k}$
are too tiny, of course, there will be no hope to produce and
detect heavy Majorana neutrinos and test the seeasw mechanism at
the LHC \cite{XingZZ09}.

\section{The minimal seesaw scenario}

The exact seesaw relation in Eq. (7) allows us to determine
$M^{}_k$ in terms of $m^{}_i$ and the mixing parameters of $V$ and
$R$. To illustrate this point, let us focus on the minimal type-I
seesaw scenario which contains only two heavy Majorana neutrinos
\cite{MSS}. In this simpler case, it is easy to obtain two real
and linear equations of $M^{}_1$ and $M^{}_2$ from Eq. (8)
%%%%%%%%%%%%%%%%%%%%%%%%%%%%%%%%%%%%
\footnote{Similar equations of $M^{}_1$ and $M^{}_2$ can also be
obtained from the exact seesaw relation in Eq. (7). A detailed
analytical calculation and numerical analysis of the dependence
and consequences of such equations will be presented elsewhere
\cite{Wu}.}:
%%%%%%%%%%%%%%%%%%%%%%%%%%%%%%%%%%%%
\begin{eqnarray}
\sum^2_{k=1} {\rm Re} R^2_{ek} M^{}_k & \;=\; & -\sum^3_{i=1} {\rm
Re}V^2_{ei} m^{}_i \; , \nonumber \\
\sum^2_{k=1} {\rm Im} R^2_{ek} M^{}_k & \;=\; & -\sum^3_{i=1} {\rm
Im}V^2_{ei} m^{}_i \; .
%     (15)
\end{eqnarray}
Note that either $m^{}_1 =0$ or $m^{}_3 =0$ must hold in the
minimal type-I seesaw model \cite{MSS}, and thus the non-vanishing
neutrino masses can be determined from current experimental data
on two independent neutrino mass-squared differences $\Delta
m^2_{21} \equiv m^2_2 - m^2_1$ and $\Delta m^2_{32} \equiv m^2_3 -
m^2_2$ corresponding to solar and atmospheric neutrino
oscillations. After a simple calculation, we arrive at
\begin{eqnarray}
M^{}_1 & \;=\; & +\frac{\displaystyle {\rm Re}R^2_{e2}
\sum^3_{i=1} {\rm Im} V^2_{ei} m^{}_i - {\rm Im} R^2_{e2}
\sum^3_{i=1} {\rm Re} V^2_{ei} m^{}_i}{\displaystyle {\rm
Re}R^2_{e1} {\rm Im}R^2_{e2} -
{\rm Im}R^2_{e1} {\rm Re}R^2_{e2}} \; , \nonumber \\
M^{}_2 & \;=\; & -\frac{\displaystyle {\rm Re}R^2_{e1}
\sum^3_{i=1} {\rm Im} V^2_{ei} m^{}_i - {\rm Im} R^2_{e1}
\sum^3_{i=1} {\rm Re} V^2_{ei} m^{}_i}{\displaystyle {\rm
Re}R^2_{e1} {\rm Im}R^2_{e2} - {\rm Im}R^2_{e1} {\rm Re}R^2_{e2}}
\; .
%     (16)
\end{eqnarray}
Using the exact and convenient parametrization of $V$ and $R$
advocated in Ref. \cite{Xing08}, we have
\begin{eqnarray}
V^{}_{e1} & \;=\; & c^{}_{12} c^{}_{13} c^{}_{14} c^{}_{15} \; ,
~~~~ V^{}_{e2} \; =\; s^{}_{12} c^{}_{13} c^{}_{14} c^{}_{15}
e^{-i\delta^{}_{12}} \; , ~~~~ V^{}_{e3} \; =\; s^{}_{13}
c^{}_{14} c^{}_{15} e^{-i\delta^{}_{13}} \; , \nonumber \\
R^{}_{e1} & \;=\; & s^{}_{14} c^{}_{15} e^{-i\delta^{}_{14}} \; ,
~~~~ R^{}_{e2} \; =\; s^{}_{15} e^{-i\delta^{}_{15}} \;
%     (17)
\end{eqnarray}
for the minimal type-I seesaw scenario, where $c^{}_{1i} \equiv
\cos\theta^{}_{1i}$ and $s^{}_{1i} \equiv \sin\theta^{}_{1i}$ (for
$i=2,\cdots,5$). There are at least two phenomenological merits of
this parametrization for our present discussions: (1) it can
automatically reproduce the standard (unitary) parametrization of
the light neutrino mixing matrix \cite{PDG08} when the non-unitary
mixing angles of $R$ are switched off; and (2) it can lead to a
very simple result of $\langle m\rangle^{}_{ee}$, which is equal
to the standard (unitary) expression of $\langle m\rangle^{}_{ee}$
multiplied by a factor $c^2_{14} c^2_{15}$ \cite{Xing08}.
Substituting Eq. (17) into Eq. (16), we obtain the explicit
results of $M^{}_1$ and $M^{}_2$ for two different patterns of the
light neutrino mass spectrum.

\subsection{Normal hierarchy ($m^{}_1 = 0$)}

In this case, it is straightforward to obtain $m^{}_2 =
\sqrt{\Delta m^2_{21}} \approx 8.8 \times 10^{-3}$ eV and $m^{}_3
= \sqrt{\Delta m^2_{21} + |\Delta m^2_{32}|} \approx 5.0 \times
10^{-2}$ eV, where we have typically input $\Delta m^2_{21} = 7.7
\times 10^{-5} ~{\rm eV}^2$ and $|\Delta m^2_{32}| = 2.4 \times
10^{-3} ~ {\rm eV}^2$ \cite{Fogli}. The expressions of $M^{}_1$,
$M^{}_2$ and $\langle m\rangle^{}_{ee}$ are given by
\begin{eqnarray}
M^{}_1 & \;=\; & - \frac{m^{}_2 s^2_{12} c^2_{13} \sin
\left(\phi^{}_2 + \phi\right) + m^{}_3 s^2_{13} \sin
\left(\phi^{}_2 - \phi\right)}{\sin \left(\phi^{}_2 -
\phi^{}_1\right)} \cdot \frac{c^2_{14}}{s^2_{14}} \; , \nonumber
\\
M^{}_2 & \;=\; & + \frac{m^{}_2 s^2_{12} c^2_{13} \sin
\left(\phi^{}_1 + \phi\right) + m^{}_3 s^2_{13} \sin
\left(\phi^{}_1 - \phi\right)}{\sin \left(\phi^{}_2 -
\phi^{}_1\right)} \cdot \frac{c^2_{14}c^2_{15}}{s^2_{15}} \; ,
%     (18)
\end{eqnarray}
together with
\begin{equation}
\langle m\rangle^{}_{ee} \; =\; c^2_{14} c^2_{15} \sqrt{m^2_2
s^4_{12} c^4_{13} + m^2_3 s^4_{13} + 2 m^{}_2 m^{}_3 s^2_{12}
c^2_{13} s^2_{13} \cos 2\phi} \;\; ,
%     (19)
\end{equation}
where $\phi \equiv \delta^{}_{13} - \delta^{}_{12}$, $\phi^{}_1
\equiv 2\delta^{}_{14} - \left(\delta^{}_{12} +
\delta^{}_{13}\right)$ and $\phi^{}_2 \equiv 2\delta^{}_{15} -
\left(\delta^{}_{12} + \delta^{}_{13}\right)$. Note that $M^{}_1
>0 $ and $M^{}_2 >0$ require that three CP-violating
phases $\phi$, $\phi^{}_1$ and $\phi^{}_2$ should not all be
vanishing; instead, they must take proper and nontrivial values.

\vspace{0.4cm}

The above results can be simplified in the limit of $s^{}_{13}
\rightarrow 0$, which actually has no conflict with current
experimental data:
\begin{eqnarray}
M^{}_1 & \;=\; & -\frac{\langle m\rangle^{}_{ee}}{s^2_{14}
c^2_{15}} \cdot \frac{\sin \left(\phi^{}_2 + \phi\right)}{\sin
\left(\phi^{}_2 - \phi^{}_1\right)} \; , \nonumber \\
M^{}_2 & \;=\; & +\frac{\langle m\rangle^{}_{ee}}{s^2_{15}} \cdot
\frac{\sin \left(\phi^{}_1 + \phi\right)}{\sin \left(\phi^{}_2 -
\phi^{}_1\right)} \; ,
%     (20)
\end{eqnarray}
and $\langle m\rangle^{}_{ee} = s^2_{12} c^2_{14} c^2_{15}
m^{}_2$. We have the following observations: (a) the magnitude of
$\langle m\rangle^{}_{ee}$ is of ${\cal O}(10^{-3})$ eV \cite{2B},
and thus it is experimentally inaccessible in the near future; (b)
the signs of $\sin (\phi^{}_2 + \phi)$ and $\sin (\phi^{}_2 -
\phi^{}_1)$ must be different, while the signs of $\sin (\phi^{}_1
+ \phi)$ and $\sin (\phi^{}_2 - \phi^{}_1)$ must be the same; and
(c) it will be possible, at least in principle, to obtain the
lower bounds on $M^{}_1$ and $M^{}_2$ if some constraints on the
non-unitary mixing angles and CP-violating phases become available
at low energies. To achieve $M^{}_1 \sim M^{}_2 \gtrsim 10^2$ GeV,
for example, Eq. (20) implies that $s^2_{14} c^2_{15} \sin
(\phi^{}_2 - \phi^{}_1) \sim s^2_{15} \sin (\phi^{}_2 - \phi^{}_1)
\lesssim 10^{-14}$ must hold. Current experimental bounds can only
provide us with $s^2_{14} + s^2_{15} \lesssim 1.1 \times 10^{-2}$
\cite{Antusch}. Hence there is a long way to go before we can
constrain $s^{}_{14}$, $s^{}_{15}$ and even the relevant
CP-violating phases to a much better degree of accuracy.

\subsection{Inverted hierarchy ($m^{}_3 = 0$)}

In this case, it is easy to obtain $m^{}_1 = \sqrt{|\Delta
m^2_{32}| - \Delta m^2_{21}} \approx 4.8 \times 10^{-2}$ eV and
$m^{}_2 = \sqrt{|\Delta m^2_{32}|} \approx 4.9 \times 10^{-2}$ eV,
where $\Delta m^2_{21} = 7.7 \times 10^{-5} ~{\rm eV}^2$ and
$|\Delta m^2_{32}| = 2.4 \times 10^{-3} ~ {\rm eV}^2$ \cite{Fogli}
have typically been input. After a straightforward calculation,
the explicit expressions of $M^{}_1$, $M^{}_2$ and $\langle
m\rangle^{}_{ee}$ are given as follows:
\begin{eqnarray}
M^{}_1 & \;=\; & - \frac{m^{}_1 c^2_{12} \sin \left(\phi^\prime_2
- \phi^\prime\right) + m^{}_2 s^2_{12} \sin \left(\phi^\prime_2 +
\phi^\prime\right)}{\sin \left(\phi^\prime_2 -
\phi^\prime_1\right)} \cdot \frac{c^2_{13}c^2_{14}}{s^2_{14}} \; ,
\nonumber \\
M^{}_2 & \;=\; & + \frac{m^{}_1 c^2_{12} \sin \left(\phi^\prime_1
- \phi^\prime\right) + m^{}_2 s^2_{12} \sin \left(\phi^\prime_1 +
\phi^\prime\right)}{\sin \left(\phi^\prime_2 -
\phi^\prime_1\right)} \cdot
\frac{c^2_{13}c^2_{14}c^2_{15}}{s^2_{15}} \; ,
%     (21)
\end{eqnarray}
and
\begin{equation}
\langle m\rangle^{}_{ee} \; =\; c^2_{13} c^2_{14} c^2_{15}
\sqrt{m^2_1 c^4_{12} + m^2_2 s^4_{12} + 2 m^{}_1 m^{}_2 c^2_{12}
s^2_{12} \cos 2\phi^\prime} \;\; ,
%     (22)
\end{equation}
where $\phi^\prime \equiv - \delta^{}_{12}$, $\phi^\prime_1 \equiv
2\delta^{}_{14} - \delta^{}_{12}$ and $\phi^\prime_2 \equiv
2\delta^{}_{15} - \delta^{}_{12}$. The requirement of $M^{}_1
>0 $ and $M^{}_2 >0$ implies that three CP-violating
phases $\phi^\prime$, $\phi^\prime_1$ and $\phi^\prime_2$ cannot
all be vanishing; instead, they must take proper and nontrivial
values.

\vspace{0.4cm}

One can see that $M^{}_1$, $M^{}_2$ and $\langle m\rangle^{}_{ee}$
are independent of the small mixing angle $\theta^{}_{13}$. Given
$\theta^{}_{12} \approx 34^\circ$ \cite{Fogli}, for example, the
maximal value of $\langle m\rangle^{}_{ee}$ in Eq. (22) is
achievable at $\phi^\prime \approx 0$; i.e.,  $\langle
m\rangle^{}_{ee}$ can maximally amount to $(m^{}_1 c^2_{12} +
m^{}_2 s^2_{12}) \approx 4.9 \times 10^{-2}$ eV \cite{2B}, which
is experimentally accessible in the near future. Again, we may
optimistically argue that it is in principle possible to obtain
the lower bounds on $M^{}_1$ and $M^{}_2$ if some constraints on
the non-unitary mixing angles and CP-violating phases become
available at low energies.

\vspace{0.4cm}

It is worth remarking that the non-unitarity of $V$ is signified
by non-vanishing $R$, whose mixing angles and CP-violating phases
are quite possible to have some nontrivial observable effects. For
example, an appreciable CP-violating asymmetry up to the percent
level is expected to show up between $\nu^{}_\mu \rightarrow
\nu^{}_\tau$ and $\overline{\nu}^{}_\mu \rightarrow
\overline{\nu}^{}_\tau$ oscillations in some medium- or
long-baseline experiments at a neutrino factory \cite{Yasuda},
just as a consequence of non-vanishing $R$. A neutrino telescope
could also be a useful tool to probe the non-unitary effect in
ultrahigh-energy cosmic neutrino oscillations \cite{Zhou}.

\section{Summary}

We have carefully examined the contributions of both light
Majorana neutrinos $\nu^{}_i$ with masses $m^{}_i$ (for $i=1,2,3$)
and heavy Majorana neutrinos $N^{}_k$ with masses $M^{}_k$ (for
$k=1,\cdots,n$) to the $0\nu\beta\beta$ decay in the type-I seesaw
mechanism, in which the light neutrino mixing matrix $V$ is
non-unitary due to the non-vanishing coupling matrix $R$ between
$N^{}_k$ and charged leptons. The exact seesaw relation allows us
to establish a straightforward relationship between $(m^{}_i,
V^{}_{\alpha i})$ and $(M^{}_k, R^{}_{\alpha k})$. We have pointed
out that the constraint $|\sum R^2_{ek} M^{-1}_k| < 5 \times
10^{-8} ~ {\rm GeV}^{-1}$ used in some literature is in most cases
too loose for a type-I seesaw mechanism either at a
superhigh-energy scale or at the electroweak or TeV scale, because
the contribution of $\nu^{}_i$ to the $0\nu\beta\beta$ decay is in
most cases dominant over the contribution of $N^{}_k$ to the same
process. Such an observation leads us to a much stronger bound on
$M^{}_k$ and $R^{}_{ek}$; i.e., $|\sum R^2_{e k} M^{}_k | < 0.23 ~
{\rm eV}$ (or $< 0.85$ eV) at the $2\sigma$ level, extracted from
the present experimental upper bound on the $0\nu\beta\beta$
decay.

\vspace{0.4cm}

We have also looked at whether the future measurements of lepton
number violation and non-unitarity of neutrino flavor mixing at
low energies are possible to shed light on the masses of heavy
Majorana neutrinos. Taking the minimal type-I seesaw scenario for
example, we have illustrated the possibility of determining or
constraining two heavy Majorana neutrino masses by using more
accurate low-energy data on the $0\nu\beta\beta$ decay and
non-unitary neutrino mixing and CP violation. Such an analysis can
simply be extended to the more general cases of the type-I seesaw
mechanism with three or more heavy Majorana neutrinos.

\vspace{0.4cm}

As stressed in Ref. \cite{Xing08}, testing the unitarity of the
light neutrino mixing matrix $V$ in neutrino oscillations and
searching for the signatures of heavy Majorana neutrinos $N^{}_k$
at TeV-scale colliders can be complementary to each other, both
qualitatively and quantitatively, in order to deeply understand
the intrinsic properties of Majorana particles. We optimistically
expect that some experimental breakthrough in this aspect will
pave the way towards the true theory of neutrino mass generation
and flavor mixing.

\vspace{1cm}

I am indebted to M. Chaichian for warm hospitality during my
visiting stay in Helsinki, where this work was started, and to T.
Ohlsson for warm hospitality during the Nordita scientific program
``Astroparticle physics --- A Pathfinder to New Physics" in
Stockholm, where this work was finalized. I am also grateful to W.
Chao, H. Zhang and S. Zhou for useful discussions and comments.
This research was supported in part by the National Natural
Science Foundation of China under grant No. 10425522 and No.
10875131.

%%%%%%%%%%%%%%%%%%%%%%%%%%%%%%%%%%%%%%%%%%%
\begin{figure*}[t]
\centering \vspace{2cm}
\includegraphics[width=110mm]{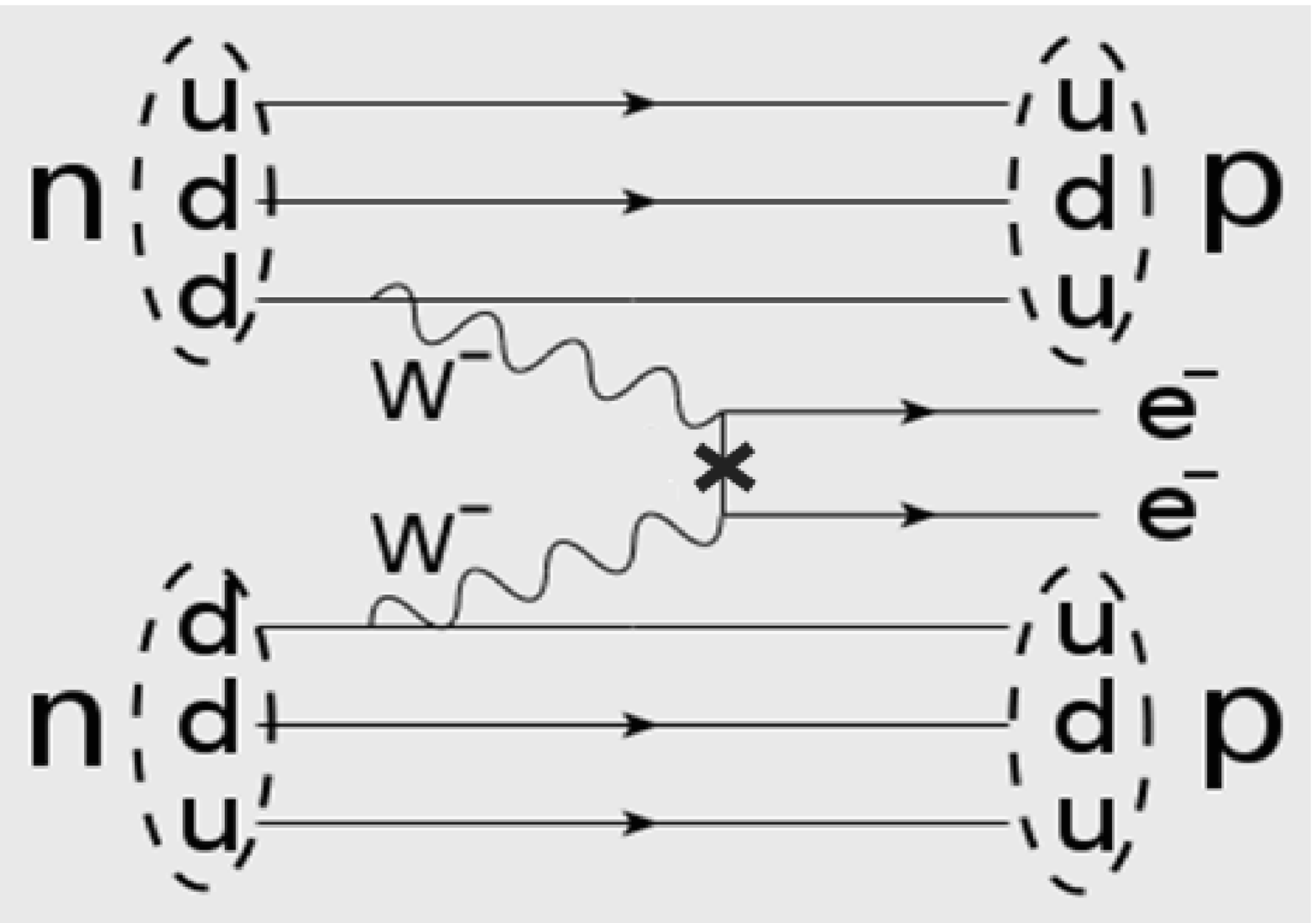}
\vspace{0.7cm} \caption{A schematic Feynman diagram for the
lepton-number-violating $0\nu\beta\beta$ decay, in which
``$\times$" stands for either light Majorana neutrinos $\nu^{}_i$
(for $i=1,2,3$) or heavy Majorana neutrinos $N^{}_k$ (for
$k=1,\cdots,n$) in the type-I seesaw mechanism.}
\end{figure*}
%%%%%%%%%%%%%%%%%%%%%%%%%%%%%%%%%%%%%%%%%%%


\begin{thebibliography}{99}
\bibitem{SNO} SNO Collaboration, Q.R. Ahmad {\it et al.},
Phys. Rev. Lett. {\bf 89}, 011301 (2002).

\bibitem{SK} For a review, see: C.K. Jung {\it et al.},
Ann. Rev. Nucl. Part. Sci. {\bf 51}, 451 (2001).

\bibitem{KM} KamLAND Collaboration, K. Eguchi {\it et al.},
Phys. Rev. Lett. {\bf 90}, 021802 (2003).

\bibitem{K2K} K2K Collaboration, M.H. Ahn {\it et al.},
Phys. Rev. Lett. {\bf 90}, 041801 (2003).

\bibitem{SS1} P. Minkowski, Phys. Lett. B {\bf 67}, 421 (1977);
T. Yanagida, in {\it Proceedings of the Workshop on Unified Theory
and the Baryon Number of the Universe}, edited by O. Sawada and A.
Sugamoto (KEK, Tsukuba, 1979), p. 95; M. Gell-Mann, P. Ramond, and
R. Slansky, in {\it Supergravity}, edited by P. van Nieuwenhuizen
and D. Freedman (North Holland, Amsterdam, 1979), p. 315; S.L.
Glashow, in {\it Quarks and Leptons}, edited by M. L$\acute{\rm
e}$vy {\it et al.} (Plenum, New York, 1980), p. 707; R.N.
Mohapatra and G. Senjanovic, Phys. Rev. Lett. {\bf 44}, 912
(1980).

\bibitem{DG} See, e.g., A. de Gouv$\rm\hat{e}$a, Phys. Rev. D {\bf 72},
033005 (2005); arXiv:0706.1732; A. de Gouv$\rm\hat{e}$a, J.
Jenkins, and N. Vasudevan, Phys. Rev. D {\bf 75}, 013003 (2007).

\bibitem{ICHEP08} For a brief review with many recent references,
see: Z.Z. Xing, Int. J. Mod. Phys. A {\bf 23}, 4255 (2008);
Plenary talk given at ICHEP2008, August 2008, Philadelphia, USA;
arXiv:0905.3903; Z.Z. Xing and S. Zhou, arXiv:0906.1757.

\bibitem{Review} For the latest review with extensive references,
see: F.T. Avignone III, S.R. Elliot, and J. Engel, Rev. Mod. Phys.
{\bf 80}, 481 (2008).

\bibitem{Seesaw} For two latest comprehensive works with extensive references,
see: F. del Aguila and J.A. Aguilar-Saavedra, Nucl. Phys. B {\bf
813}, 22 (2009); A. Atre, T. Han, S. Pascoli, and B. Zhang, JHEP
{\bf 0905}, 030 (2009).

\bibitem{CC} CUORICINO Collaboration, C. Arnaboldi {\it et
al.}, Phys. Rev. C {\bf 78}, 035502 (2008).

\bibitem{Fogli} G.L. Fogli {\it et al.}, Phys. Rev. D {\bf 78},
033010 (2008).

\bibitem{MSS} P.H. Frampton, S.L. Glashow, and T. Yanagida,
Phys. Lett. B {\bf 548}, 119 (2002). For a review with extensive
references, see: W.L. Guo, Z.Z. Xing, and S. Zhou, Int. J. Mod.
Phys. E {\bf 16}, 1 (2007).

\bibitem{Xing08} Z.Z. Xing, Phys. Lett. B {\bf 660}, 515 (2008).

\bibitem{Referee} W.C. Haxton and J. Stephenson, Prog. Part. Nucl. Phys. {\bf 12},
409 (1984).

\bibitem{SSReview} For some recent reviews with extensive references, see:
H. Fritzsch and Z.Z. Xing, Prog. Part. Nucl. Phys. {\bf 45}, 1
(2000); Altarelli and F. Feruglio, New J. Phys. {\bf 6}, 106
(2004); Z.Z. Xing, Int. J. Mod. Phys. A {\bf 19}, 1 (2004); S.F.
King, Rept. Prog. Phys. {\bf 67}, 107 (2004); R.N. Mohapatra and
A.Yu. Smirnov, Ann. Rev. Nucl. Part. Sci. {\bf 56}, 569 (2006); A.
Strumia and F. Vissani, hep-ph/0606054.

\bibitem{London} G. B$\acute{e}$langer, F. Boudjema, D. London,
and H. Nadeau, Phys. Rev. D {\bf 53}, 6292 (1996).

\bibitem{Antusch} S. Antusch, C. Biggio, E. Fernandez-Martinez, M.B.
Gavela, and J. L$\rm\acute{o}$pez-Pav$\rm\acute{o}$n, JHEP {\bf
0610}, 084 (2006); A. Abada, C. Biggio, F. Bonnet, M.B. Gavela,
and T. Hambye, JHEP {\bf 0712}, 061 (2007).

\bibitem{Minkowski} See, e.g., J. Bernabeu {\it et al.},
{\it Phys. Lett. B} {\bf 187}, 303 (1987); W. Buchmuller and D.
Wyler, {\it Phys. Lett. B} {\bf 249}, 458 (1990); W. Buchmuller and
C. Greub, {\it Nucl. Phys. B} {\bf 363}, 345 (1991); A. Pilaftsis,
{\it Z. Phys. C} {\bf 55}, 275 (1992); A. Datta and A. Pilaftsis,
{\it Phys. Lett. B} {\bf 278}, 162 (1992); G. Ingelman and J.
Rathsman, {\it Z. Phys. C} {\bf 60}, 243 (1993); C. A. Heusch and P.
Minkowski, {\it Nucl. Phys. B} {\bf 416}, 3 (1994); D. Tommasini, G.
Barenboim, J. Bernabeu, and C. Jarlskog, {\it Nucl. Phys. B} {\bf
444}, 451 (1995); J. Gluza, {\it Acta Phys. Polon. B} {\bf 33}, 1735
(2002); J. Kersten and A. Yu. Smirnov, {\it Phys. Rev. D} {\bf 76},
073005 (2007); W. Chao, S. Luo, Z.Z. Xing, and S. Zhou, Phys. Rev. D
{\bf 77}, 016001 (2008).

\bibitem{XingZZ09} Z.Z. Xing, arXiv:0905.3903; and references
therein.

\bibitem{Wu} X.G. Wu and Z.Z. Xing, work in progress.

\bibitem{PDG08} Particle Data Group, C. Amsler {\it et al.},
Phys. Lett. B {\bf 667}, 1 (2008).

\bibitem{2B} S.M. Bilenky {\it et al.}, Phys. Lett. B {\bf 465},
193 (1999); Z.Z. Xing, Phys. Rev. D {\bf 65}, 077302 (2002); Phys.
Rev. D {\bf 68}, 053002 (2003); Phys. Lett. B {\bf 618}, 141
(2005); S. Pascoli, S.T. Petcov, and W. Rodejohann, Phys. Lett. B
{\bf 558}, 141 (2003).

\bibitem{Yasuda} E. Fernandez-Martinez, M.B. Gavela, J.
L$\rm\acute{o}$pez-Pav$\rm\acute{o}$n, and O. Yasuda, Phys. Lett.
B {\bf 649}, 427 (2007); Z.Z. Xing, Phys. Lett. B {\bf 660}, 515
(2008); S. Luo, Phys. Rev. D {\bf 78}, 016006 (2008); S. Goswami
and T. Ota, Phys. Rev. D {\bf 78}, 033012 (2008); G. Altarelli and
D. Meloni, Nucl. Phys. B {\bf 809}, 158 (2009); Z.Z. Xing,
arXiv:0901.0209; M. Malinsky, T. Ohlsson, and H. Zhang,
arXiv:0903.1961; S. Antusch, M. Blennow, E. Fernandez-Martinez,
and J. L$\rm\acute{o}$pez-Pav$\rm\acute{o}$n, arxiv:0903.3986; W.
Rodejohann, arXiv:0903.4590; M. Malinsky, T. Ohlsson, Z.Z. Xing,
and H. Zhang, arXiv:0905.2889.

\bibitem{Zhou} Z.Z. Xing and S. Zhou, Phys. Lett. B {\bf 666}, 166
(2008); Z.Z. Xing,  Nucl. Instrum. Meth. A {\bf 602}, 58 (2009).
\end{thebibliography}
\end{document}